# MODELLING VARIABILITY FOR SYSTEM FAMILIES

***Shamim Hasnat Ripon and Kamrul Hasan Talukder***
Computer Science and Engineering Discipline
Khulna University, Khulna 9208, Bangladesh
email: sh_ripon@yahoo.com

***Md. Khademul Islam Molla***
Department of Computer Science and Technology
University of Rajshahi, Rajshahi 6205, Bangladesh
email: khadem_ru@yahoo.com

***ABSTRACT***

*In this paper, an approach to facilitate the treatment with variabilities in system families is presented by explicitly modelling variants. The proposed method of managing variability consists of a variant part, which models variants and a decision table to depict the customisation decision regarding each variant. We have found that it is easy to implement and has advantage over other methods. We present this model as an integral part of modelling system families.*

## 1.0 INTRODUCTION

Designing, maintaining and developing a good software system is a major challenge still in this 21st century. Reusing [1] of existing good solutions or techniques is currently a promoting solution approached by researchers, whereas reusing is not always a better offer when this is done in the code level.

In [2], a definition of a software product line is given as "a set of software-intensive systems sharing a common, managed set of features that satisfy the specific needs of a particular market segment or mission and that are developed from a common set of core assets in a prescribed way".

Presently, product line technology is a way of improving the software development lifecycle and reuse by providing facilities to reuse the model of the system family. By reusing rather than recreating the work products of the system families, it is possible to increase the productivity and decrease the possible errors significantly [2].

The main idea of software product line is to explicitly identify all the activities which are common to all members of the family as well as which are different and then arrange them in a model. This implies a huge model which will help the stakeholders to be able to trace any design choices and variability decisions as well. Finally, the derivation of the product is done by selecting the required variants from the model and configuring them according to product requirements.

The system family approach adopts ideas from domain engineering [2] which comprises three parts. Firstly, the domain analysis forms the commonality and variability data basis. Following this, the domain design phase forms the flexible generic architecture. Finally, based on this architecture, an application is derived in the implementation phase.

Today, most of the efforts in product line development are relating to software architecture [3], detailed design and code. Our work focuses on the variability issues in the domain modelling phase.

Our particular interest is to model variabilities in system families. The commonalities found across the system families are easy to handle as they are simply integrated into the generic architecture and a part of every family member. However, intricate problems arise for the variabilities found across members. There is a need to take proper treatment for these variabilities.

In this paper, we describe our approach to facilitate the treatment with variabilities in system families by explicitly modelling variants. The model comprises all the information concerning the variants, and according to our approach, this model will be an indivisible part of modelling system families. We present our approach using a very simple example of *Hall Booking System*, a system used both academically and commercially.

The remaining part of the paper is organised as follows. Section 2 describes about variability and system family. Related works are presented in Section 3. A brief introduction of the Hall Booking System is described in Section 4. Section 5 describes the proposed variability model and its customisation process is illustrated in Section 6. In Section 7, we give a brief discussion of the proposed model, and following this, Section 8 concludes with the summary of the work and future direction.

## 2.0 VARIABILITY AND SYSTEM FAMILY

An explicit variability model as a carrier of all variability related information like specifications, interdependencies, origins, etc., can play an important and maybe the central role in successful variability management.

In developing a system family, a variability model is to be created in the domain engineering phase which scopes the system families and develops the means to rapidly produce the members of the family. It serves two distinct but related purposes. First, it can record decisions about the product as a whole including identifying the variants for each member of the product line. Second, it can support the application engineering phase by providing proper information and mechanism for the required variant during product generation (Fig. 1).

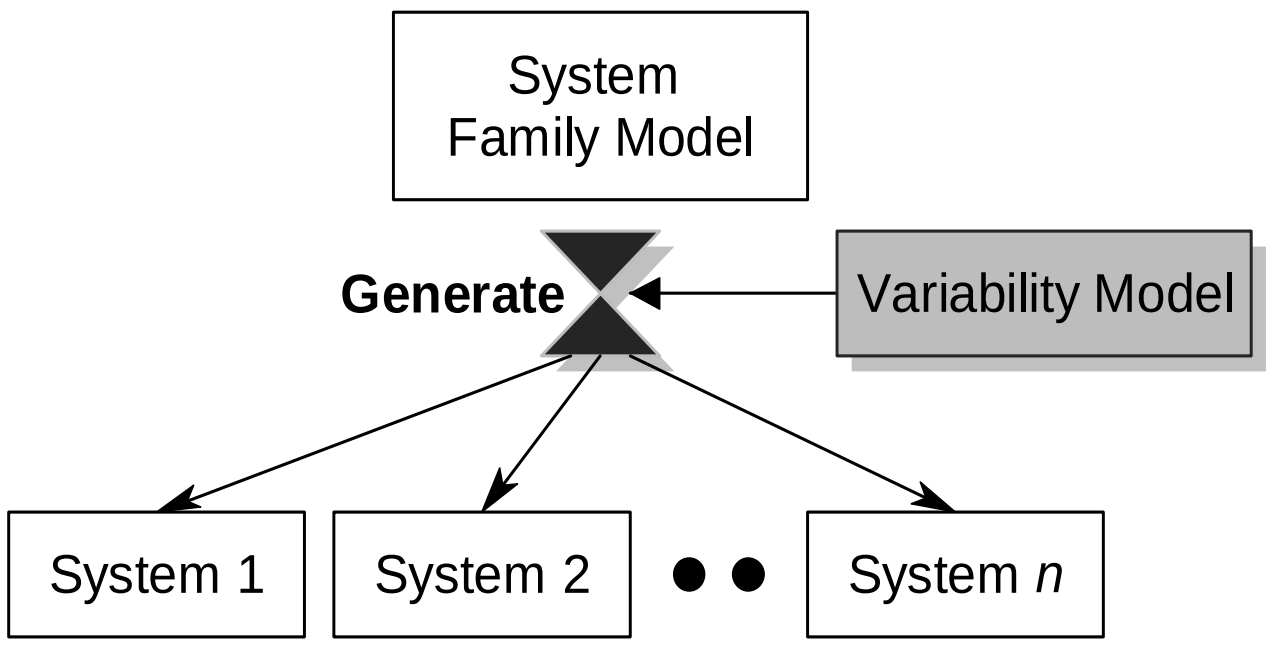

Fig. 1 : Systems derived from System Family model using Variability Model

Though it is possible to create a software system family without developing a distinct variability model where the domain is well understood by all members (i.e. developer, stakeholders), the variability model offers some potential advantages:

- It provides a distinct document characterising all the variants of the family as a whole for both domain engineers and stakeholders interested in the product line.
- It has a place for recording the selection and application related decision of each variant which is part of a family and provides support to implement any member product.
- The required variant of any particular product can be validated and verified by it.

While developing the variability model, we have some objectives to be met by the model. The model should specify both the commonalities and variants of the family members. In the case of variants, it should contain their application areas, constraints, possible values, dependencies, etc. Along with these objectives, the variability model should contain the customisation and configuration structure of each variant of the system family which will guide the application engineer to generate any product from the system family.

## 3.0 RELATED WORK

Modelling variability resembles with feature modelling as both the approaches model the variabilities found among system family members. Feature modelling is described in FODA [4]. In this method, features are modelled hierarchically in a graphical form which classifies the features as mandatory, optional or alternative. Some features can also be classified as "OR" features [5]. A coupling of feature and variability can be found in [6] where variabilities are represented throughout the development process with features.

In Reuse driven Software Engineering Business (RSEB) approach [1], the UML notations are extended with variation points to cater for variant requirements and a generic software model is customised by attaching one or several variants to its variation points.

Several other methods use UML for modelling system families with some extension of UML to represent the variabilities [7, 8]. In this paper, we will use some of these extensions. Object oriented method is integrated with feature oriented method in FeatuRSEB [9] which extended the UML based RSEB method with feature model.

Most of the above mentioned methods use the feature diagram of FODA for representing the variabilities of a product line. Whereas the feature diagram has some restrictions of variability to specify features to some binding times as well as decomposition types. There are some attributes needed to choose a variant which are absent in the feature diagram like availability sites (i.e. when, where and whom for a feature is available), variability mechanisms, binding models (static or dynamic), binding occurrences, description, etc. There are several approaches for solving variability at the code level like using macros, templates, meta-programming techniques [5] and frame technology [10]. However, in the domain modelling of product line, there is a lack of approaches for modelling variabilities.

In our approach, we use some simple UML extension to model the system family and we propose a variability model for system families.

## 4.0 HALL BOOKING SYSTEM OVERVIEW

We use *Hall Booking System* family to illustrate our variability modelling mechanism. The system is used in academic institutions to reserve tutorial rooms and lecture halls, at companies to reserve meeting rooms, and at hotels to reserve rooms and conference facilities, etc. In another sense, the system can be used for either academic or non-academic purposes. Users can manage their own reservation with the system. The main purpose and the core functionality are similar across the *Hall Booking System* family; however, there are many variants on the basic theme. One of the basic variants is the charging of booking system. Whenever the system is used for academic purposes, no charge is needed for booking halls, whereas there may be a need to charge for booking halls in other areas. In some systems, there are facilities available for seasonal booking as well as multiple bookings.

The descriptive part of the hall booking system consists of feature diagrams, domain defaults modelled in UML and domain defaults instrumented with variants in UML. Domain defaults describe a typical system in a domain. Our *Hall Booking System* default models cover the functionalities shown in Fig. 2.

Make reservation
Delete reservation
Modify reservation
Search/Retrieve reservation
Add a resource (Hall)
Delete a resource (Hall)
Modify a resource
Search/receive a Hall

Fig. 2: Functionalities covered by default Hall Booking System

Feature models are often used to depict the different features of a system. A part of the features of Hall Booking System is shown in Fig. 3. Extensions of feature diagram described in [5] have been used here. Mandatory features appear in all the members of the family whereas variant features appear in some members of the family. Variant features are also classified as Optional, Alternative and Or features. An example of optional feature is *Reservation Charge* option. An alternative feature describes one of many features. An example of alternative feature is *Reservation Mode* which can be either *Single* or *Block*. An or-feature describes any of many features. For example a *Block Reservation* can be made by multiple rooms or multiple times or by both. Variants may depend on other variants.

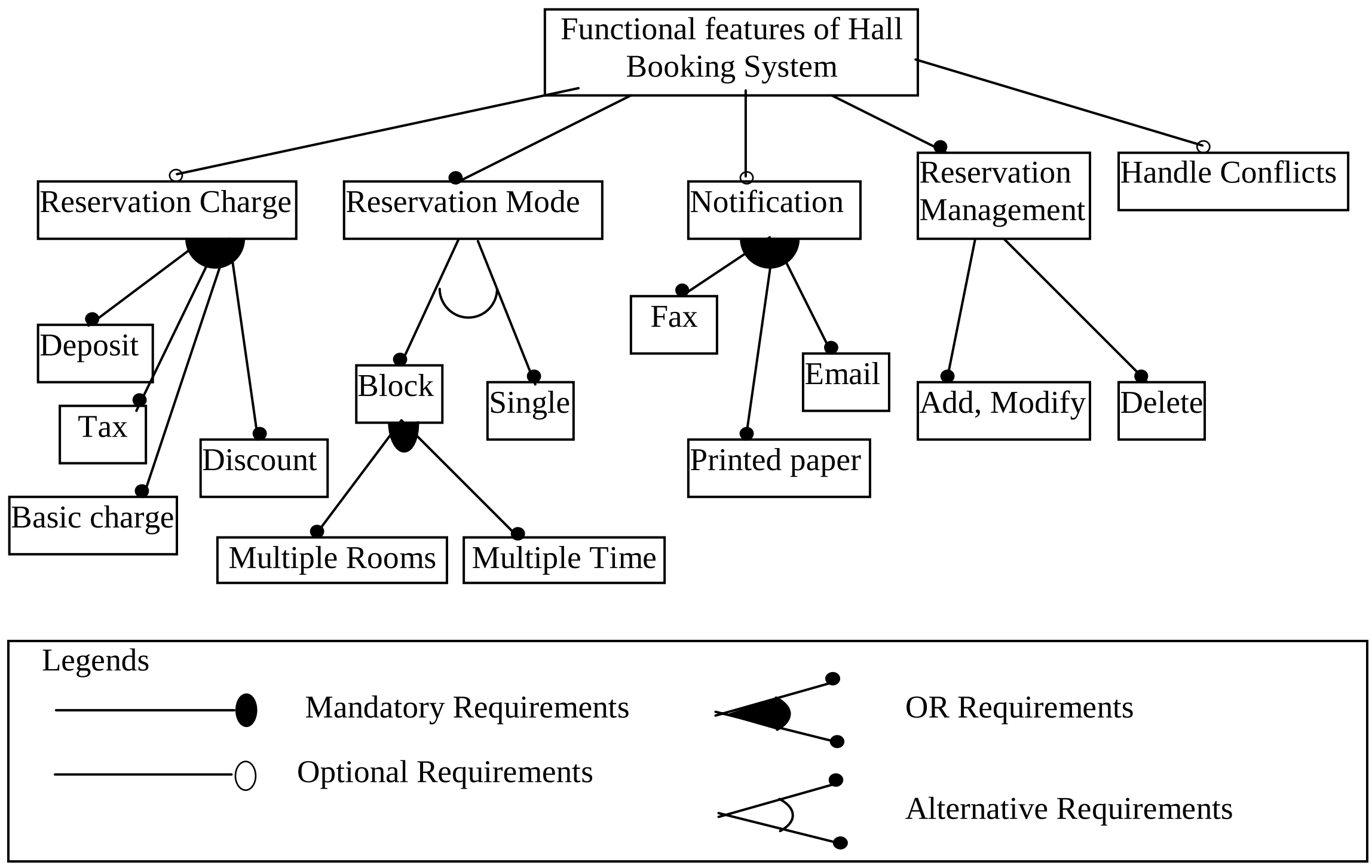

Fig. 3: Partial feature diagram of Hall Booking System

## 5.0 THE PROPOSED VARIABILITY MODEL

Our proposed solution consists of a variant model and a decision table, upon which a model of a member of the system family can be developed. The variability model is shown in Fig. 4.

### 5.1 Model Organisation

The left part of the Fig. 4 depicts the system family model. The system family model consists of a Default Model and a Variant part. The commonalities found across the members of the system family are identified and they are modelled as Default. This will be the main part which will remain in all the products of the system family, and hence, they are called default part. These defaults are modelled using UML diagrams. These UML models will help the developer to develop the common part of the product easily. After comparing different family members, the variant parts are also identified. The default parts as well as the variant parts can also be represented using feature diagrams which hierarchically describe the features of the system.

After identifying the default part and the variant part, models are drawn by combining all the variants to the default part as it will give a clear view of the whole system family to both the developers and the stakeholders, and this will help during the selection of product variants.

The feature model as described earlier depicts the features hierarchically and is also used to graphically represent the features of the system family as well as the variant model. However, all the variant related information cannot be represented by it. We use it only to show the output of the variability model.

When there is a need to generate a product from the system family then requirements are captured from the stakeholders. These requirements consist of the default part of the family and some selected parts from the variants. During product generation, these requirements are checked against the variability model.

The main focus of Fig. 4 is the 'Variability Model' which consists of a Variant Model and a Decision Table. The Variant Model contains the configuration and application information of each variant. Every variant has its own application area and configurability. They have dependencies on other variants and sometimes, their selection depends on some special criteria of the required product. There is also a need of traceability of each variant with the model elements to identify which variants are responsible for modelling which part of the family. Along with these variant properties, there are some basic properties of variants like each variant has some possible values and these values have some relationships among them and they are considered during the selection of these values. The variant model contains all these information.

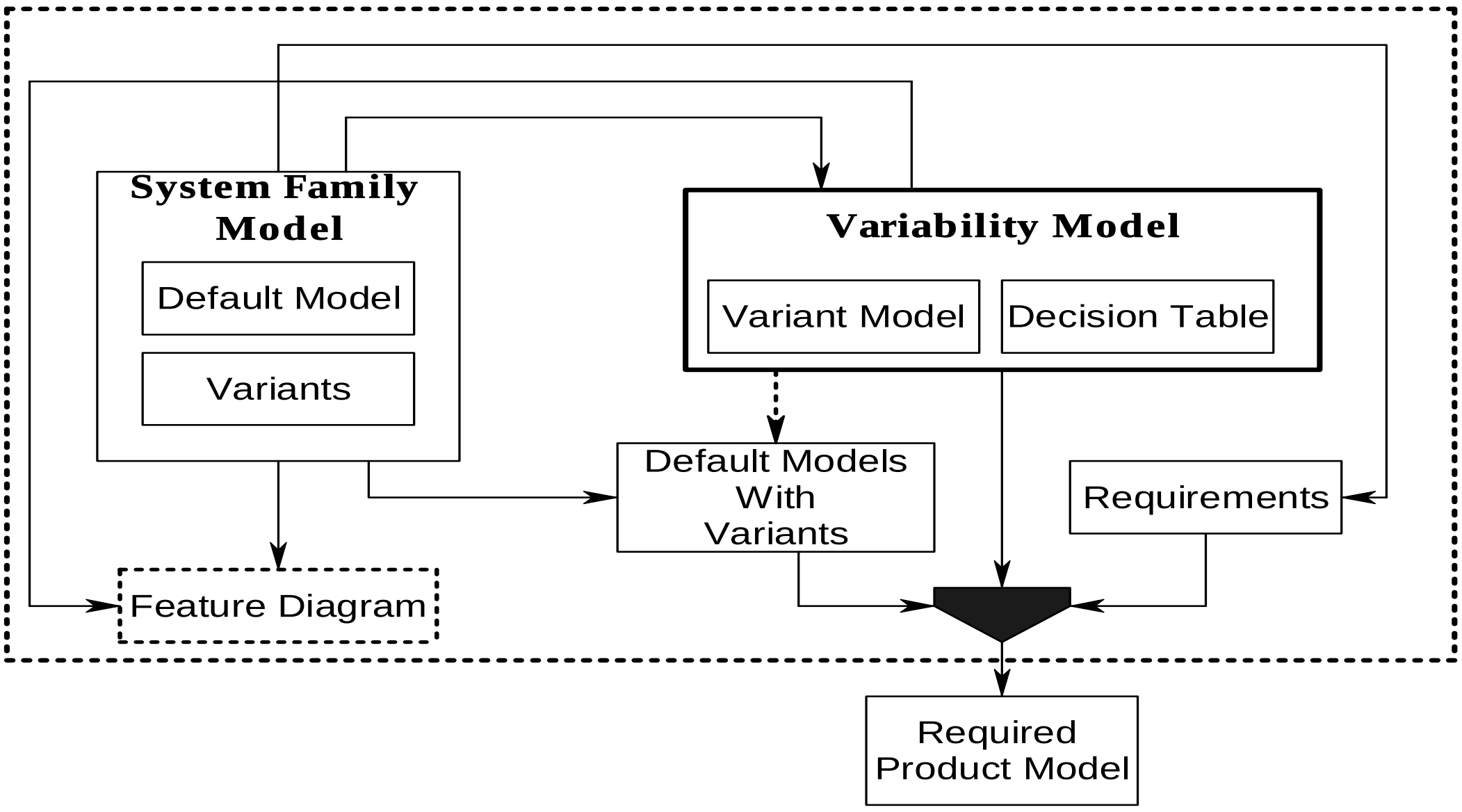

Fig. 4: Proposed Variability Model with System Family model

A decision table is then created for the system families. This table represents the set of choices which differentiate the family members of the system family. The decision table captures those decisions which an application engineer must make to define a member of the system family, such as the decision table captures the variant dependencies among the variants, e.g., if the selection of variant Y is dependent on selection of any values of variant X. The decision table guides the derivation of any family member from the system family.

The variant model that we have created here can be represented using the usual textual notation arranged in hierarchical order or it can be represented in a tabular format where each row represents a variant and the columns represent the individual properties such as values, dependencies, applicable areas, etc. Both the textual and tabular approaches have equal significance. We are using XML (Extensible Markup Language) to represent the variant model which will help in future to construct a tool for developing the system family.

### 5.2 Applying the Model

We have experimentally validated our approach by applying the Variability Model to a portion of the product line: *Hall Booking System*. This system can be used in both academic and commercial purposes. UML diagrams are used to represent the model of the system family. Simple extension mechanisms of UML, namely stereotypes and tagged values, are used here. The stereotype <<variant>> designates a model element as a variant and the tagged values are used to keep trace of the models and their corresponding variant elements. The activity diagram in Fig. 5 shows the steps to reserve a hall of *Hall Booking System*, which combines both the default elements as well as the variants. The tagged values in the UML models are pointing to the corresponding entry in the variant model as well as decision table which keep traces of each modelling element, e.g. in Fig. 5, the tagged value of the variant *Notification* is given *V.4* which means that the variant has corresponding entity in the variant model and in the decision table whose number is *V.4*. However, adding these types of stereotypes and tagged values to all modelling elements possibly results in a complex model which will be difficult to understand and maintain.

Tabular representation has been used in this paper to represent the variability model. The rows in the table represent the variants and their properties are represented in the columns. Any number of columns can be added to the table according the number of the properties required to represent the variants. Whenever there is a new variant, it is just needed to add another entry at the end of the table to place the variant in the variability model. Every variant is uniquely identified by their number which keeps traceability of the variant. If a variant has multiple choices, then they are considered as values of the variant and they are given the numbers followed by the variant which identify them clearly.

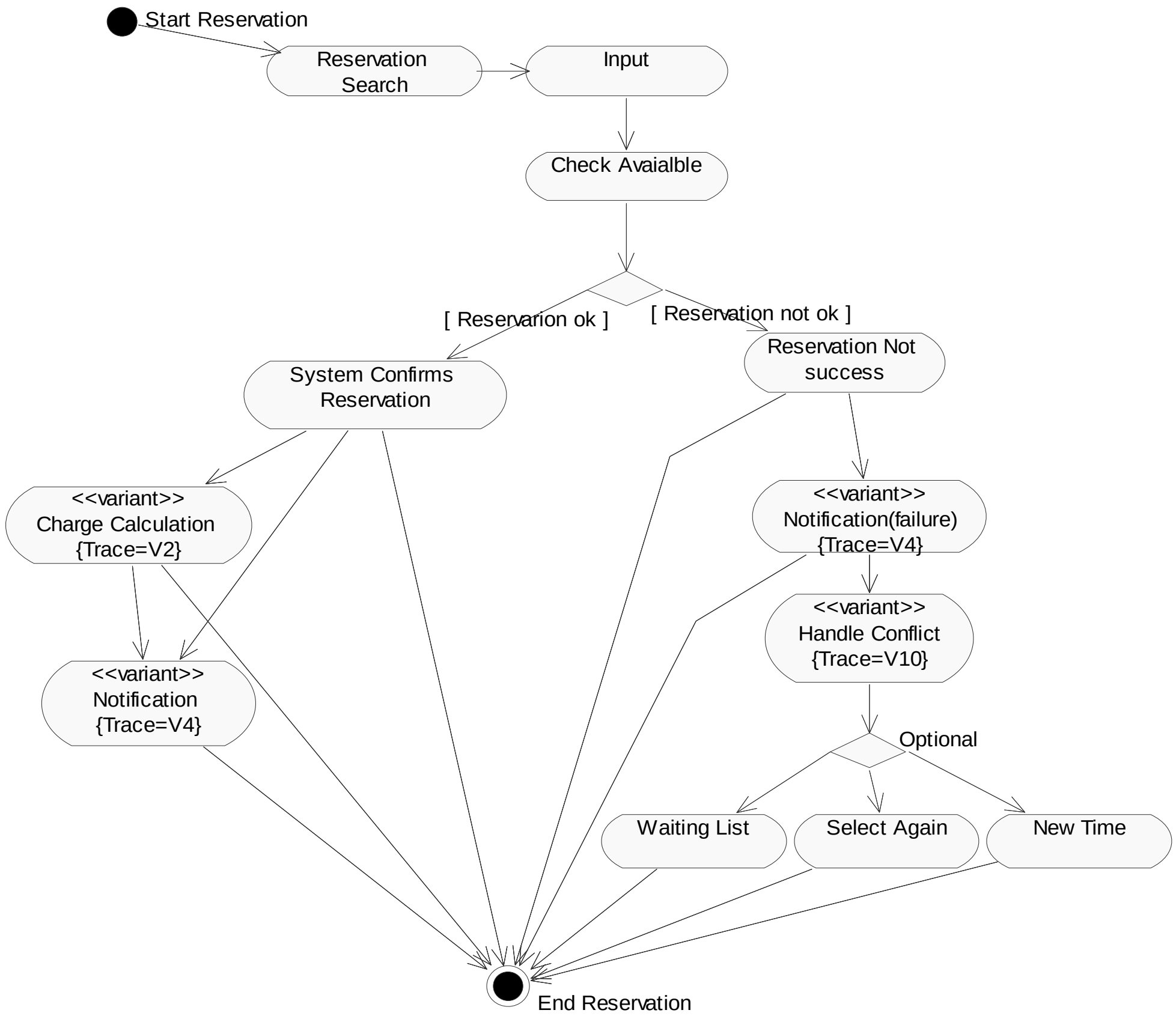

Fig. 5: Activity diagram of reserving a hall of Hall Booking System

The *OR* and *Alternative* relations of variant values show the relationships among them and they have their usual meaning like in the feature diagram (as shown in Fig. 6). *Applicable Area* denotes the particular areas applicable for any variant. The *Hall Booking System* has two major applicable areas for the variants which are *Academic* and *Non Academic*. When a variant is applicable to both areas, then it is mentioned as *ALL*. The dependency is identified by the numbers given to the variants or their values.

| Variant | Values of variant | Relations | Applicable Area | Dependency |
|---|---|---|---|---|
| V1. Reservation Mode | V1.1 Single<br>V1.2 Block | Alternative | All | None |
| V2. Reservation Charge | V2.1 Deposit<br>V2.2 Tax<br>V2.3 Discount<br>V2.4 Refund | OR | Non Academic | None |
| V3. Block Reservation | V3.1 Multiple Room<br>V3.2 Multiple time | OR | All | V1.2 |
| V4. Notification | V4.1 Fax<br>V4.2 Email<br>V4.3 Printed Paper | OR | All | None |
| V5. Reservation Discount | V5.1 Block Discount<br>V5.2 Seasonal discount | OR | Non Academic | V2.3, V1.2 |

Fig. 6: Variant Model for Hall Booking System (Partial)

A decision table is then derived from the variant model. A small part of the decision table is given in Fig. 7. In the table, each variant is placed in a column. Its description, possible choices and traces to the variant are shown in other columns. When inclusion or exclusion of any variant depends on a variant or its values then that variant is placed in the subordinate position showing the values for which it can be chosen. For example, *Reservation Mode* variant has two possible values either *Single* or *Block*. So these two values are placed in the subordinate position of the parent variant (Reservation Mode) in the decision table (Fig. 7). Similarly when *Block reservation* is to be considered then it needs to check its values (Multiple Room and Multiple Time) and these values along with their other information will be also be subordinated. For each variant and for each of their values, the decision table can have a corresponding entity.

| Variant | Description | | Values | Traces |
|---|---|---|---|---|
| Reservation Mode | What is the reservation mode? | | Single, Block | V1 |
| Block reservation | Block | What is the type of block reservation? | Multiple Room, Multiple Time | V3 |
| Reservation charge | Single | How is the charge for reservation? | Deposit, Tax, Discount | V2 |

Fig. 7: Decision Table for Hall Booking System (Partial)

So, after getting requirements from stakeholders, those are checked with the variant model and the decision table. The variability model will guide the application engineer to properly choose the required variants very easily. Therefore, the overall product generation process will be faster and less erroneous.

## 6.0 CUSTOMISATION PROCESS

By using the proposed variability model, it is now possible to derive any member product from the system family model according to the user's requirements. In the proposed variability model, all the possible variants are added to the default models, which help the developer to select the proper variants according to the user's requirements. Suppose there is a need for a *Hall Booking System* for the university purpose where the user wants to be notified by printed paper and there is no need to handle conflicts when a reservation is not available. When the developer gets the above requirements, the required variants are extracted from the variant model. As the required system is for the academic purpose, those variants which are only for non-academic purposes like reservation charge, are discarded. Using the variant dependency information given in the variant model, other dependant variants are also extracted along with the required variants. So the developer does not need to check all the dependencies of the system family. This results a smaller variant model as shown in Fig. 8.

| Variant | Values of variant | Relations | Applicable Area | Dependency |
|---|---|---|---|---|
| V1.<br>Reservation Mode | V1.1 Single<br>V1.2 Block | Alternative | All | None |
| V3.<br>Block Reservation | V3.1 Multiple Room<br>V3.2 Multiple time | OR | All | V1.2 |
| V4.<br>Notification | V4.3 Printed Paper | | All | None |

Fig. 8: Variant model for customised Hall Booking System (Partial)

Similarly, the decision table is also reduced containing only the decisions for the required features. By using the decision table, the user can take decisions on variants. Suppose the user wants to reserve a hall for multiple time slots. Using the dependency information of the decision table and the variant model, it is clear that if this variant is required then there is a need to choose the reservation mode to *Block*. Following this way the developer can select the proper variants according to the requirements and consequently can derive the proper product model. A customised UML activity diagram is shown in Fig. 9. By using this customised model, the developer can now develop the required customised product in a suitable developing language.

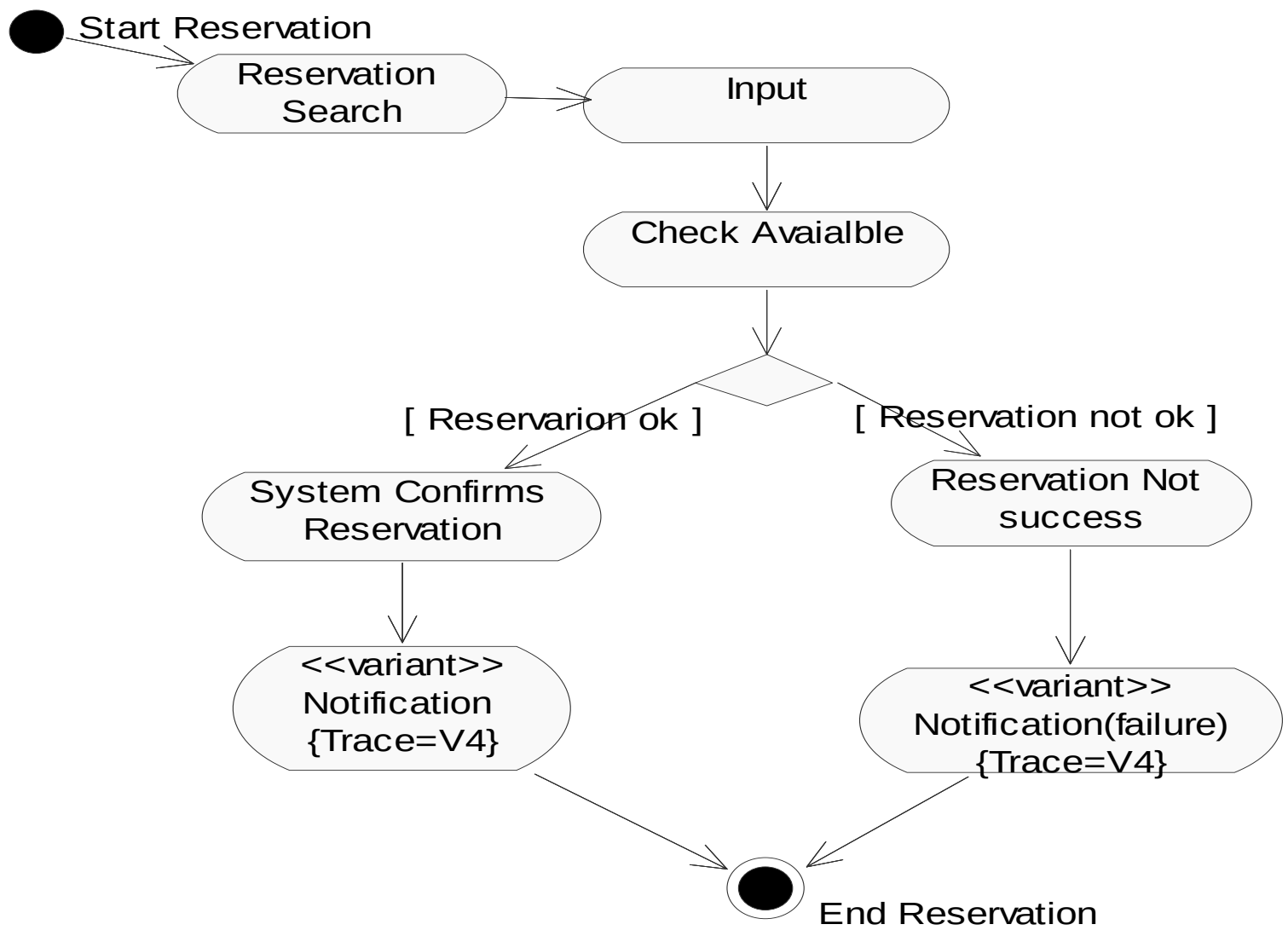

Fig. 9: Customised Activity diagram of reserving a hall

## 7.0 DISCUSSION

The proposed variability model that we have described in this paper offers some features which are useful in handling variants of system family. However, the experiment we have done to explain the model has some limitations.

Some of the important and valuable features of the variability model are listed below:

- The proposed method contains two parts; the variant model is one of them. The variant model grasps the variant related information very easily when all the possible variants of the family are available, whereas when any new variant appears for any members, it can be easily included in the model without intervening other variants. Similarly, unwanted variants can be removed from the model.

- After developing the variant model, it is very easy to locate any variant and its related dependency information just by looking the corresponding table entry without going through every possible entry in the model.

- During application engineering process, it is possible to extract a smaller model from the whole variant model for the required product by selecting only those variants which have the required application area and this is done by looking only at the corresponding entry in the model. So having a smaller model, the application engineer can generate the required application very easily. This also helps in creating a smaller decision table for the variants.

- The decision table derived from the variant model helps the application engineer take the variant related decision for selecting any variant for the product he/she is generating.

- The variability model is presented in a text based format which helps better in any kind of processing of the variants than that of graphical representation.

We believe that our approach has the potential to reduce the complexity in modelling the variabilities of system families. At the same time, our approach and scope of the experiment have the following limitations:

- *Experiment on larger scale*
  We have experimented with selective views of the system family and on a small-sized scale. We plan to cover a wider spectrum of modelling variants on a larger scale in future. We believe that the proposed model can be applied to other views of system families such as state-transition diagram, object-collaboration diagram, etc.

- *Address complex variant dependencies*
  So far, we have been dealing with relatively simple functional variant dependencies. We have yet to extend research to non functional variants. Also, other system families may give rise to different types of dependencies (such as time-based dependencies) that will require specialised approach.

## 8.0 CONCLUSIONS

Successful development of software system families requires appropriate organisation and management of the products involved. A significant characteristic of developing system families is the management of variabilities, which is a crucial factor for the success of system family approach.

Most of the system family development approaches do not focus on the variant related information. In this paper, we presented an innovative approach for modelling variants based on the ideas of existing approaches. UML models have been used to model the system family with their simple extension mechanisms. We propose the variant modelling approach as an integral part of developing a system family. This model helps the application engineers in implementing any product from the system family by providing a systematic representation of the variants.

Much work remains to be done. Currently, we are using a XML based prototype for representing the variability model and we are focusing on creating a tool to support the overall modelling process in an automotive and effective way.

## BIOGRAPHY

**Shamim Hasnat Ripon** obtained his B.Sc in Computer Science and Engineering from Khulna University, Bangladesh in September 1997. He joined as a lecturer at Computer Science and Engineering Discipline, Khulna University after completing his B.Sc. His research areas include Software Engineering, Software Product Line and Software Reuse.

**Kamrul Hasan Talukder** completed his BSc in Computer Science and Engineering (CSE) from Khulna University, Bangladesh in December 1999. He has been a lecturer in CSE Discipline of Khulna University since May 2000. His research interest mainly includes Embedded Systems, Formal Verification, Software Engineering, Complier Construction and Algorithms.

**Md. Khademul Islam Molla** obtained his B.Sc and M.Sc in Electronics and Computer Science (ECS) from Shahjalal University of Science and Technology, Sylhet, Bangladesh in 1995 and 1997 respectively. After completing his Masters he joined as a lecturer at ECS Department in September 1997. Then he joined as a lecturer of the Department of Computer Science and Technology, University of Rajshahi, Bangladesh. He is presently serving as an Assistant Professor at Computer Science and Technology Department of Rajshahi University, Bangladesh. His research interest includes Speech and Language Processing, Computer Vision and Pattern Recognition, Model Verification for Embedded System Design and Software Engineering.